\documentclass[pra,a4paper,superscriptaddress,twocolumn,showpacs,amsmath,amssymb,floatfix,amstex]{revtex4}

\usepackage[dvips]{graphicx}
\usepackage{bm,pstricks,longtable}

\input{epsf}
\newcommand{\ket}[1]{| #1\rangle}                       %

\newcommand{\bra}[1]{\langle #1}                        %

\begin{document}

\title{Delocalization induced by nonlinearity in systems with disorder}
\author{Ignacio Garc\'ia-Mata}
\affiliation{\mbox{Laboratoire de Physique Th\'eorique, UMR 5152 du CNRS, 
Universit\'e  Toulouse III, 31062 Toulouse, France}}
\author{Dima L. Shepelyansky}
\homepage[]{http://www.quantware.ups-tlse.fr/dima}
\affiliation{\mbox{Laboratoire de Physique Th\'eorique, UMR 5152 du CNRS, 
Universit\'e  Toulouse III, 31062 Toulouse, France}}

\date{\today}

\begin{abstract}
We study numerically the effects of nonlinearity on
the Anderson localization in lattices with disorder
in one and two dimensions. The obtained results
show that at moderate strength of nonlinearity
a spreading over the lattice in time 
takes place with an algebraic growth of 
number of populated sites $\Delta n \propto t^{\nu}$.
This spreading continues up to a maximal dimensionless time scale
$t=10^9$ reached in the numerical simulations.
The numerical values of $\nu$ are found to be
approximately $0.15 - 0.2$ and $0.25$
for the dimension $d=1$ and 2 respectively
being in a satisfactory agreement with the theoretical 
value $d/(3d+2)$. During the computational times $t \leq 10^9$ 
the localization is preserved 
below a certain critical
value of nonlinearity. 
We also discuss the properties
of the fidelity decay induced
by a perturbation of nonlinear field.
\end{abstract}


\maketitle

\section{Introduction}
The phenomenon of the Anderson localization \cite{anderson1958}
in systems with disorder
has been extensively studied for electron
transport and linear waves (see e.g. \cite{montambaux}).
A remarkable experimental progress 
with the Bose-Einstein condensates (BEC) in optical lattices 
(see e.g. reviews \cite{pitaevskii,oberthaler,inguscio2008})
stimulated the interest to investigations of the effects of
nonlinearity on localization. 
At present the signatures of localization of BEC
in one-dimensional (1D) optical disordered lattices
have been detected by different experimental groups
\cite{inguscio2005,aspect2005,inguscio12005,ertmer2005,aspect2006}.
The effects of nonlinearity appear also 
for experiments with BEC in kicked optical lattices 
\cite{phillips2006,summy2006,steinberg}
where the quantum chaos in the 
Chirikov standard map (kicked rotator) \cite{chirikov}
is investigated. 
A similar type of problem
also comes out for propagation of nonlinear waves
in disordered photonic lattices which are now
actively studied  experimentally \cite{segev2007,silberberg2008}.
In addition to that the problem of lasing 
in random media \cite{cao}
is also linked to the interplay of 
localization and nonlinearity that makes it related to 
the important field of nonlinear wave propagation in disordered media
\cite{kivshar}. In this work we concentrate our studies on 
the time dependent wave packet spreading in presence
of disorder and nonlinearity leaving aside the 
problem of directed flow and scattering in nonlinear media
(see e.g. Refs. in \cite{kivshar} and more recent
\cite{pavloff1}).

The theoretical treatment of the interplay
between localization and nonlinearity
uses numerical simulations
(see e.g. 
\cite{dls1991,dls1993,pavloff,aspect2007,flach2008,dls2008,lebowitz})
and various analytical tools 
(see e.g. \cite{skipetrov2000,shapiro2007,wellens,shapiro2008}). 
However, even rather powerful analytical tools
\cite{skipetrov2000,shapiro2007,wellens,shapiro2008} do not allow
to obtain the full solution of this rather complex problem.
The existing rigorous mathematical
results show that for a sufficiently small
nonlinearity there exists a Kolmogorov-Arnold-Moser
integrable localized regime for 
almost all initial conditions \cite{frolich}
but these results are applicable only to unrealistically small
strength of nonlinearity. Due to that the numerical
simulations become especially important
for investigation of this problem.
For numerical studies it is especially convenient
to use a discrete lattice that allows to
push numerical simulations to extremely
large times. In addition to that  
the time evolution on a lattice
is closely linked to the problem
of energy propagation in complex
molecules, e.g. proteins, where nonlinear couplings
give transitions between localized linear modes
\cite{james2005,flach2008,lebowitz}. 
One of the examples of such a nonlinear oscillator
chain is the Frenkel-Kontorova model
in the pinned phase where the linear sound
modes are localized in space \cite{zhirov}.

Recently, the interplay of the Anderson localization and
nonlinearity has been investigated by a number of mathematical methods
where a certain number of interesting mathematical results
has been obtained \cite{soffer,wang1,wang2}. However, 
these methods still should be developed further  to understand 
the asymptotic properties of spreading in 
the lattice at moderate strength of nonlinearity.

In this paper we further develop 
the old \cite{dls1991,dls1993} and recent studies \cite{dls2008} of 
the discrete Anderson nonlinear Schr\"odinger equation (DANSE)
and present large scale numerical simulations
of this model in one and two dimensions d (1D, 2D).
In addition we perform numerical 
simulations for the kicked nonlinear rotator model
(KNR) introduced in \cite{dls1993}.
Our numerical results obtained on
dimensionless time scales up to $t =10^9$
show that at moderate nonlinearity,
above a certain threshold,
the wave packet spreads unlimitedly over
the lattice in such a way that the 
squared displacement of the packet on the lattice
grows according to the algebraic law
$R^2 \propto t^\alpha$
with the exponent $\alpha \approx 0.3 - 0.4$
for $d=1$ and $\alpha \approx 0.25$ for $d=2$.
This dependence is in a satisfactory
agreement with the 1D estimates \cite{dls1993} 
which give $\alpha=2/5$ and 
the analytical estimates of this paper 
which give $\alpha=1/4$ for 2D. We also
study the fidelity decay which
shows interesting properties for the nonlinear
evolution described by our model.

The paper is composed as follows: in Section II
we give the model description and present simple 
estimates; the results for 1D and 2D are presented in 
Sections III and IV respectively, the properties
of nonlinear fidelity decay are discussed in Section IV; 
the conclusions are  given in Section V.

\section{Model description and analytical estimates}
Our system is described by the DANSE model: 
\begin{equation}
i \hbar{{\partial {\psi}_{\boldsymbol{n}}} \over {\partial {t}}}
=E_{\boldsymbol{n}}{\psi}_{\boldsymbol{n}}
+{\beta}{\mid{\psi_{\boldsymbol{n}}}\mid}^2 \psi_{\boldsymbol{n}}
 +V ({\psi_{\boldsymbol{n+1}}}+ {\psi_{\boldsymbol{n-1}})}\;,
\label{eq1}
\end{equation}
where 
$\beta$ characterizes nonlinearity,
$V$ is a hopping matrix element
on nearby sites, on-site
energies are randomly and homogeneous distributed in the range
$-W/2 < E_n < W/2$, and the total probability is normalized to unity
$\sum_{\boldsymbol{n}} {\mid{\psi_{\boldsymbol{n}}}\mid}^2 =1$.
Here, $\boldsymbol{n}$ is the lattice index, in 1D it is
an integer, in 2D it is an integer vector
of lattice indexes $\boldsymbol{n}=(n_x,n_y)$. 
For $\beta=0$ and weak disorder
all eigenstates are exponentially localized
with the localization length $l \approx 96 (V/W)^2 $ (1D)
at the center of the energy band 
and $\ln l \sim (V/W)^2$ in 2D
\cite{kramer1993}. 
Hereafter we set for convenience $\hbar=V=1$, 
thus the energy coincides with the frequency. 
We emphasize here that the DANSE (\ref{eq1})
exactly describes recent experiments with one-dimensional 
disordered waveguide lattices (cf. Eq. (1) in \cite{silberberg2008}),
and it also serves as a paradigmatic model for a wide class of physical 
problems where interplay of nonlinearity and disorder is important. 
The DANSE can be considered as the Gross-Pitaevskii equation (GPE)
\cite{pitaevskii} taken on a discretized lattice.
In 1D this model was studied recently in \cite{flach2008,dls2008}.

To understand the evolution properties of system (\ref{eq1})
it is convenient to expand $\psi_{\boldsymbol{n}}$
in the basis of localized eigenmodes at $\beta=0$ \cite{dls1993}:
$A_{\boldsymbol{n}}=
\sum_{\boldsymbol{m}}Q_{{\boldsymbol{n}},{\boldsymbol{m}}} C_{\boldsymbol{m}}$
where $A$ and $C$ are amplitudes in the basis of sites
and eigenmodes respectively. Due to the localization of linear
eigenmodes with length $l$ we have for the transformation matrix
$Q_{{\boldsymbol{n}}{\boldsymbol{m}}}
\sim l^{-d/2}\exp(-|\boldsymbol{n}-\boldsymbol{m}|/l - i\chi_{{\boldsymbol{n}},{\boldsymbol{m}}})$, where $\chi$ are some random phases.
From (\ref{eq1}) it follows that the amplitudes 
$C$ in the linear eigenbasis are described by the equation
\begin{equation}
i {{\partial C_{\boldsymbol{m}}} \over {\partial {t}}}
=\epsilon_{\boldsymbol{m}} C_{\boldsymbol{m}}
+ \beta \sum_{{\boldsymbol{m_1}}{\boldsymbol{m_2}}{\boldsymbol{m_3}}}
V_{{\boldsymbol{m}}{\boldsymbol{m_1}}{\boldsymbol{m_2}}{\boldsymbol{m_3}}}
C_{\boldsymbol{m_1}}C^*_{\boldsymbol{m_2}}C_{\boldsymbol{m_3}}
\label{eq2}
\end{equation}
where $\epsilon_{\boldsymbol{m}}$
are the eigenmode energies. 
The transitions between linear eigenmodes
appear only due to the nonlinear $\beta$-term
and the transition matrix elements are
$V_{{\boldsymbol{m}}{\boldsymbol{m_1}}{\boldsymbol{m_2}}{\boldsymbol{m_3}}}=
\sum_{\boldsymbol{n}}Q^{-1}_{{\boldsymbol{n}}{\boldsymbol{m}}}
Q_{{\boldsymbol{n}}{\boldsymbol{m_1}}} 
Q^{*}_{{\boldsymbol{n}}{\boldsymbol{m_2}}} Q_{{\boldsymbol{n}}{\boldsymbol{m_3}}}
\sim 1/l^{3d/2}$ \cite{dls1993}. There are about $l^{3d}$ random 
terms in the sum in (\ref{eq2})
with $V \sim l^{-3d/2}$ so that we have
$id C/dt \sim \beta C^3$.  
We assume that the probability is distributed over 
$\Delta n > l^d$ states of the lattice basis. 
Then from the normalization
condition we have $C_{\boldsymbol{m}} \sim 1/(\Delta n)^{1/2}$
and the transition rate to new 
non-populated states in the basis $\boldsymbol{m}$
is $\Gamma \sim \beta^2 |C|^6 \sim \beta^2/(\Delta n)^3$.
Due to localization these transitions take place on a size $l$
and hence the diffusion rate in the 
distance $\Delta R \sim (\Delta n)^{1/d}$ of $d-$~dimensional 
$\boldsymbol{m}-$~space is 
$d (\Delta R)^2/dt \sim l^2 \Gamma \sim \beta^2 l^2 /(\Delta n)^3
\sim \beta^2 l^2 /(\Delta R)^{3d} $. At large time scales 
$\Delta R \sim R$ and we obtain
\begin{equation}
\Delta n \sim R^d \sim (\beta l)^{2d/(3d+2)} t^{d/(3d+2)} \; .
\label{eq3}
\end{equation}
Thus as in \cite{dls1993} for $d=1$ we have $R^2 \propto t^{2/5}$
and for $d=2$ this gives $R^2 \propto t^{1/4}$, while for large
$d$ the scaling is independent of $d$: $\Delta n \propto t^{1/3}$.

The relation (\ref{eq3}) assumes that the 
dynamics of nonlinear chain (\ref{eq1}) is chaotic.
On a first glance it seems that it cannot be the case
since as soon as $\Delta n$ grows with time
the nonlinear frequency shift 
$\delta \omega \sim \beta |\psi_n|^2 \sim \beta/\Delta n$
decreases. However, the physical importance relays  
not on the shift value itself but on its ratio
to a frequency spacing between 
frequencies of excited modes which is $\Delta \omega \sim 1/\Delta n$.
The latter relation results from the fact that all 
the frequencies are distributed in a finite energy (frequency) band
and therefore $\Delta n$ states excited inside such a band
have energy and frequency spacing $\Delta \omega \sim 1/\Delta n$.
The dynamics is chaotic if 
the overlap parameter 
$S = \delta \omega/\Delta \omega \sim \beta > \beta_c \sim 1$.
It is important to stress that $S$ is independent of $\Delta n$.
By its nature this criterion is somehow different 
from the usual Chirikov resonance-overlap criterion \cite{chirikov}
since in our case the unperturbed system is
represented by a set of linear oscillators
while in \cite{chirikov} the oscillators are nonlinear.
However, the condition $\delta \omega > \Delta \omega$
looks rather natural since in the opposite limit
 $\delta \omega \ll \Delta \omega$ the coupling
between modes is very weak if the linear frequencies
are linearly independent (that should be true in a disordered 
potential). In addition the investigations
of three nonlinear oscillators with nonlinear
couplings performed in \cite{chirikov1}
indeed confirmed the criterion $S>1$.
Therefore from the criterion $S >1$ we obtain
that above some critical nonlinear coupling
$\beta > \beta_c \sim const$
the dynamics remains chaotic
even if probability spreads over larger and larger
parts of the lattice. This spreading 
should follow the relation (\ref{eq3}).
During this process the local Lyapunov exponent
$\lambda \sim \delta \omega \sim \beta/\Delta n$
decreases to zero since the system size is unlimited
but locally the dynamics is chaotic. 

Another argument in 
favor of unlimited spreading can be obtained
on the basis of certain similarities and parallels with the
Frenkel-Kontorova chain.
In this nonlinear chain the number of
configurations static in time 
($d\psi_{\boldsymbol{n}}/dt=0$ in (\ref{eq1}))
grows exponentially with the length of the chain
while the energy splitting between these configurations
drops exponentially with the chain length (see e.g. \cite{zhirov}).
Therefore, due to this energy quasi-degeneracy between 
these static configurations,
it is rather natural to expect that during the  time evolution
a spreading over all these configurations 
continues unlimitedly. 

For $\beta > \beta_c$
this spreading corresponds to 
a regime of strong chaos with mixing of all modes.
The situation for  $\beta < \beta_c$ 
may have other mechanisms of slow chaos
with slower spreading and should be analyzed separately.
For example, the typical spacing in the resonant terms
in Eq.~(\ref{eq2}) is $\Delta_2 \sim E_m+E_{m_2}-E_{m_3}-E_{m_4}
\sim 1/l^{2d}$ and it is smaller than the coupling matrix element
$\beta V_{mm_1m_2m_3} \sim \beta/l^{3d/2}$
for $\beta l^{d/2} > 1$. Therefore, it is possible that
for $l^{-d/2} < \beta <\beta_c \sim 1$ there may be a propagation of 
two-modes-pairs on a distance much larger than $l$ in a certain
similarity with a quantum dynamics of the 
two interacting particles (TIP) 
in a random potential discussed in \cite{dls1994}.
Indeed, Eq.~(\ref{eq2}) can be viewed as a mean field
approximation for the TIP Hamiltonian considered in \cite{dls1994}.
In analogy with the TIP problem it is possible to expect that the
distribution of the probability in the basis of linear eigenmodes
$C_m$ will be characterized by the Breit-Wigner shape:
$w_{m,m_1} =|C_m(t) C_{m_1}(t)|^2 \sim 
\Gamma/[(E-\epsilon_m-\epsilon_{m_1})^2+\Gamma^2/4]$
(see e.g. discussion in \cite{dls1995} for TIP).
Here, the value of $\Gamma$ is given by the above estimates.
However, the verification of this Breit-Wigner relation 
requires further numerical tests with a projection on the eigenbasis
of linear modes that was not done in this work.
In this paper we concentrate our studies on the regime
$\beta \sim 1 > \beta_c$.

\section{Numerical results in 1D}
To test the above theoretical predictions we perform
the numerical simulations of the time evolution given  by  
Eq.~(\ref{eq1}). The split operator scheme on a time step $\Delta t$ 
is used for integration: 
\begin{equation}
\psi_{{\boldsymbol{n}},{\boldsymbol{m}}}(t+\Delta t)=
\hat{Q}\hat{V}\psi_{{\boldsymbol{n}},{\boldsymbol{m}}}(t) \; ,
\label{eq4}
\end{equation}
where $\hat{Q}=\exp(-i (E_{n,m} + \beta {\mid{\psi_{n,m}}\mid}^2) \Delta t)$
is a diagonal operator in the lattice space
and the application of the hopping operator $\hat{V}$ is done by
the fast Fourier transform to the conjugated space
where $\hat{V}$ becomes diagonal taking in 2D the form
$\hat{V}=\exp( - 2 i\Delta t (\cos \hat{\theta}_{n_x} +\cos \hat{\theta}_{n_y} ))$.
For the results presented in next Sections we used $\Delta t=0.1$
and the averaging was done over $N_d=10$ realisations of disorder.
We checked that a variation of  $\Delta t$
by a factor 2-4 does not affect 
the numerical data for short times (e.g. $t < 20$)
and on the large time scales $t \sim 10^7$
the statistical behavior of the results remains unchanged.
Of course, on large times the exact values of $\psi_n$
are different for different  $\Delta t$ due to exponential instability of 
dynamics. But the integration scheme (\ref{eq4}) 
is symplectic and preserves the total probability
exactly while the total energy is preserved  approximately
with the accuracy of $1\%$. 
Indeed, the final integration step generates high frequency
 $\omega_{int}=2\pi/\Delta t \approx 60$ that is significantly larger
than the energy band width of the linear problem.
We checked that for 1D this
integration scheme gives the same results as other schemes used
in \cite{dls2008}. The lattice size in 1D was $N=2^{11}$ site
and in 2D $N=256 \times 256$. The initial state was chosen 
with all probability on one site in the middle of the lattice. 
We note that for the KNR model (\ref{eq5})
the integration precision is on the level of double precision of the computer
since the integration is done by the fast Fourier transform from 
coordinate to momentum representation and in the each representation 
the integration is performed exactly up to the computer double precision.

\begin{figure}[htb!]
\begin{center}
\includegraphics[width=8cm]{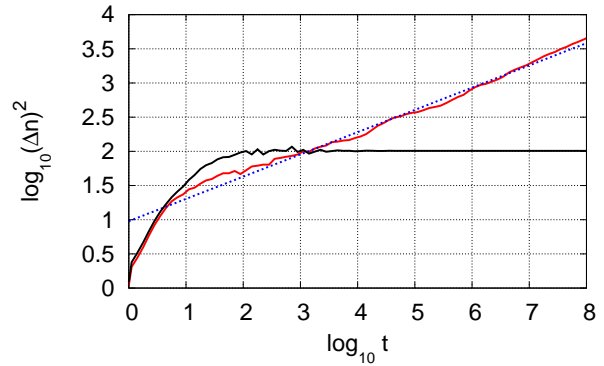} 
\end{center}
\caption{(Color online) Time dependence of the 
averaged second moment  $(\Delta n)^2$ in 1D for the 
disorder strength $W=4$ at 
$\beta=0$ (black) and $\beta=1$ (red/gray). 
The straight line shows the fit 
for $100 \leq t \leq 10^8$ with the slope
$\alpha_1=0.325 \pm 0.003$. Here and below
the logarithms are decimal.} 
\label{fig1}
\end{figure}

\begin{figure}[htb!]
\begin{center}
\includegraphics[width=8cm]{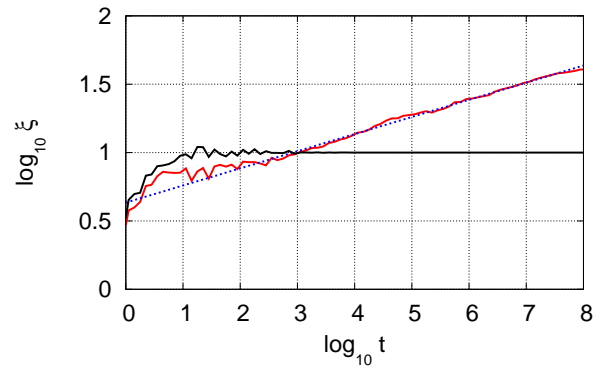} 
\end{center}
\caption{(Color online) Time dependence of the 
averaged IPR $\xi$ for the parameters of Fig.~{\ref{fig1}}
for $\beta=1$ (red/gray curve)
and $\beta=0$ (black curve).
The straight line shows the fit with the slope
$\nu=0.125  \pm 0.001$.}
\label{fig2}
\end{figure}

To characterize the properties of time evolution we 
compute one-site probability $w_{\boldsymbol{n}}=|\psi_{\boldsymbol{n}}|^2$,
second moment of the probability distribution
$(\Delta n)^2$ in 1D and $(\Delta R)^2 = (\Delta n_x)^2+(\Delta n_y)^2$
in 2D, and the inverse participation ratio (IPR)
$\xi = 1/ \sum_{\boldsymbol{n}} {w_{\boldsymbol{n}}}^2 $
which gives an effective number of sites populated by the wave packet
if all ${w_{\boldsymbol{n}}}^2$ probabilities are of the same order.
To suppress fluctuations the  quantities $(\Delta n)^2$, 
$(\Delta R)^2$, $\xi =1/\sum_{\boldsymbol{n}} {w_{\boldsymbol{n}}}^2 $  
are averaged over time intervals which are equally spaced
in $\log t$. In addition the logarithms of these quantities are
averaged over $N_d=10$ disorder realizations. The dependence on time
is fitted by the algebraic dependencies
$(\Delta n)^2 \sim t^{\alpha_1}$, $(\Delta R)^2 \sim t^{\alpha_2}$,
$\xi \sim t^\nu$ with the exponents $\alpha_1, \alpha_1, \nu$.

The numerical results presented in Figs.~\ref{fig1},\ref{fig2}
give values of exponents $\alpha_1=0.325 \pm 0.003$ and
$\nu=0.125  \pm 0.001$. The value of  $\alpha_1$ is in agreement
with the data obtained in \cite{dls2008} where
it was found that $\alpha_1=0.306 \pm 0.002$ at $W=4$.
Indeed, it should be noted that the standard
deviation $\Delta \alpha_1 \approx 0.014$ for fluctuations in 
the exponent $\alpha_1$ from one disorder realization 
to another \cite{dls2008} is larger than the 
the formal standard error of the fit 
of averaged data (with $\Delta \alpha_1 \approx 0.003$).
If all states inside the width $\Delta n$ are populated
in a homogeneous way then we should have
$(\Delta n)^2 \sim \xi^2$ and $\alpha_1=2\nu$.
The numerical data of Figs.~\ref{fig1},\ref{fig2} give
this ratio to be $\alpha_1/2\nu = 1.3$ instead of 1.
This indicates that the probability inside the width
$\Delta n$ is distributed in inhomogeneous way.
It is possible that the spreading has certain multi-fractal
properties that give deviations from usual relations
between high moments. At the same time our data
clearly confirm that the IPR grows in unlimited way
with time (see Fig.~\ref{fig2}). This is different from
the claim presented in \cite{flach2008}.
We attribute this difference to the fact that 
in \cite{flach2008} the data have been presented only for one
disorder realization and no data for fits and their 
statistical accuracy have been given. At the same time the data
of \cite{flach2008} for the second moment $(\Delta n)^2$
are consistent with the results presented here and in \cite{dls2008}.

\begin{figure}[htb!]
\begin{center}
\includegraphics[width=8cm]{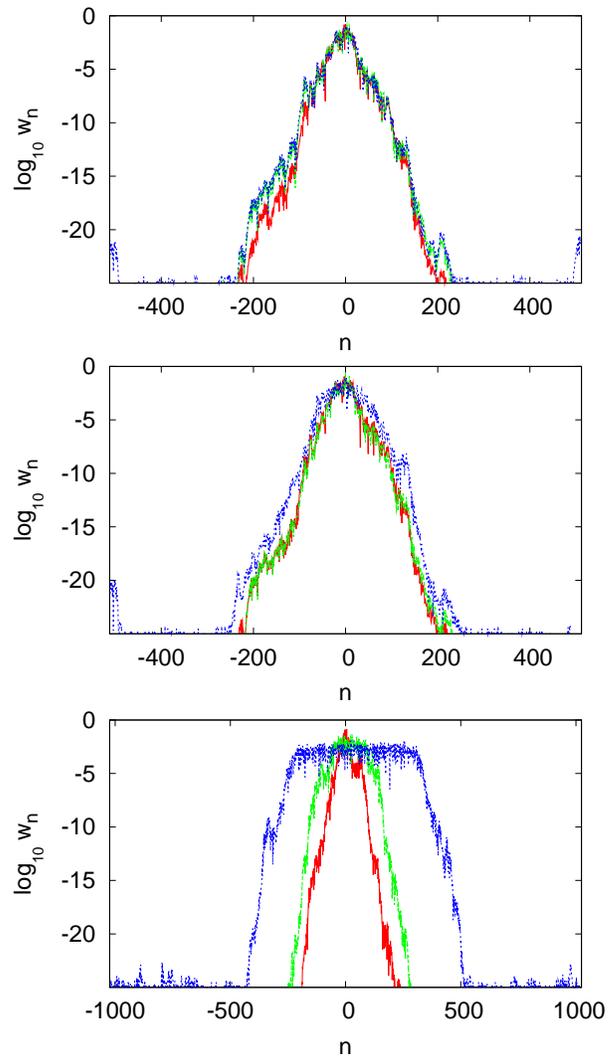} 
\end{center}
\caption{(Color online) KNR data (\ref{eq5}) 
for probability distribution
$w_n$ at times $t=10^3$ (red/gray);
$10^6$ (green/gray), $10^9$ (blue black)
for $\beta=0$ (top panel with overlapped curves) 
with basis size $N=2^{10}$, for
$\beta=0.03$ (middle panel) with the same basis size and 
for $\beta=1$ (bottom panel, curves are from bottom to top
at $|n| \approx 200$ for time
from $t=10^3$ to $10^9$) at $N=2^{11}$;
here $k=3$, $T=2$.}
 \label{fig3}
\end{figure}

\begin{figure}[htb!]
\begin{center}
\includegraphics[width=8cm]{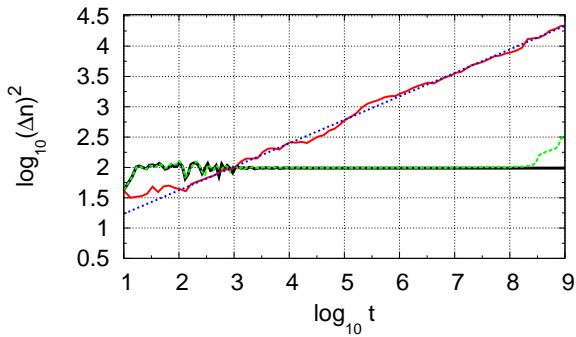} 
\end{center}
\caption{(Color online) The spreading of the second moment 
$(\Delta n)^2$ with number of map iterations $t$
for the KNR model for the parameters of Fig.~\ref{fig3}:
$\beta=0$ (black curve), $N=2^{10}$;
$\beta=0.03$ (dashed green/light gray curve), $N=2^{10}$ and 
$\beta=1$ (red/gray curve), $N=2^{11}$.
The straight line shows the fit 
for $100 \leq t \leq 10^9$ with the slope
$\alpha_1=0.387 \pm 0.003$.}
 \label{fig4}
\end{figure}

\begin{figure}[htb!]
\begin{center}
\includegraphics[width=8cm]{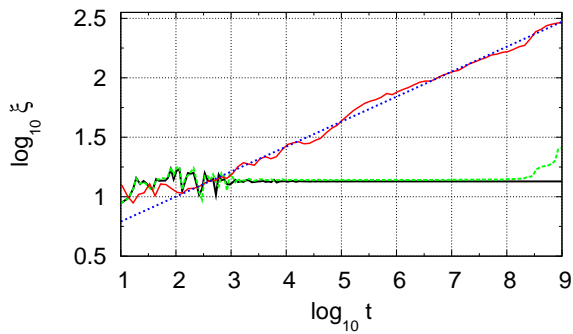} 
\end{center}
\caption{(Color online) 
Same as in Fig.~\ref{fig5} for the IPR $\xi$ in the KNR model.
The straight line shows the fit 
for $100 \leq t \leq 10^9$ with the slope
$\nu=0.210 \pm 0.002$.}
\label{fig5}
\end{figure}

To obtain more results for larger times
we also performed numerical simulations for the KNR
model introduced in \cite{dls1993}.
It time evolution is described by the map
for the wave function:
\begin{equation}
\psi_n(t+1)=e^{-iT\hat{n}^2/2-i\beta|\psi_n|^2}e^{-ik\cos \hat{\theta}}\psi_n(t) \; ,
\label{eq5}
\end{equation}
where $(\hat{n},\hat{\theta})$ are the conjugated operators
with the commutation relation $[\hat{n},\hat{\theta}]=-i$
and $\psi$ is periodic in $\theta$. For $\beta=0$ this is the 
model of kicked rotator 
where all quasienergy eigenstates are exponentially localized
with a localization length $l \approx k^2/2$
\cite{chirikov,fishman}. The propagation operator is similar
to the one of (\ref{eq4}) with $\Delta t =1$,
due to that it is possible to perform $t=10^9$
map iterations of (\ref{eq5}) for the same CPU time as for 
(\ref{eq4}). The numerical results are presented in Figs.3-5.

These results show that unlimited spreading of probability 
over the sites $n$ takes place at moderate values of $\beta \sim 1$. 
The probability distribution over $n$
has a plateau followed by exponential tails, inside the plateau
the probability is homogeneously distributed and  the width of
the plateau grows with time (see Fig.~\ref{fig3}). The second moment
of the distribution and the IPR grow algebraically in time
with the exponents $\alpha_1 = 0.387 \pm 0.003 $ 
and $\nu =0.210 \pm 0.002$ respectively (Figs.~\ref{fig4},\ref{fig5}).
In view of statistical fluctuations we consider that
these values are in a good agreement with the theory
estimates (see Eq.~(\ref{eq3}) and \cite{dls1993}). The relation
$\alpha_1=2\nu$ also works with a relatively weak deviation
from the theory. For the KNR model the agreement with the theory
is better than for the model (\ref{eq1}).
The possible reason is that in the KNR all linear eigenmodes
have the same localization length $l \approx k^2/2$
while for the DANSE the localization length depends on the energy
value inside the energy band that gives stronger
statistical fluctuations and require longer times
for the observation of the asymptotic algebraic growth.
Also, it is possible that the stronger deviations of 
the exponent values from the theory
in 1D DANSE model are related to the absence of good
diffusive approximation for the 1D Anderson model
while for the KNR model the diffusive approximation 
works rather well. 

At small values of $\beta =0.03$ the probability 
distribution remains localized during enormously long
times $t \leq 10^8$ but for larger time $t \sim 10^9$
the distribution grows slightly, also $\xi$ and $(\Delta n)^2$
are increased by a factor 2 and 3 respectively 
(see Figs.~\ref{fig3},\ref{fig4},\ref{fig5}).
It remains unclear if this is a fluctuation
or if there is a very slow (logarithmic ?) 
spreading. In any case the behavior for small
nonlinearity $\beta < \beta_c \sim 0.03$
is qualitatively different compared to the case of moderate
values of $\beta \sim 1$ being in a qualitative agreement with
the theoretical expectations.

\section{Numerical results in 2D}

\begin{figure}[ht!]
\begin{center}
\includegraphics[width=8cm]{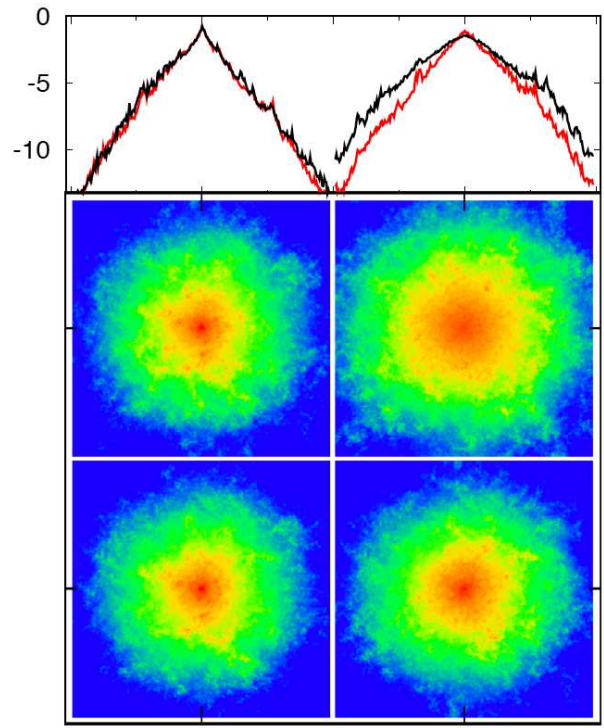} 
\end{center}
\caption{(Color online) Probability distribution
$w_{n_x,n_y}$ inside the square $256 \times 256$ at $W=10$
for $\beta=0$ (left column) and $\beta=1$ (right column)
and time $t=10^4$ (bottom panels), $10^6$ (middle panels);
probability is proportional to color with 
maximum at red/gray and zero at blue/black.
Top panels show the decimal logarithm
of the integrated probability
$w_{n_x}=\sum_{n_y} w_{n_x,n_y}$ for
$-128 \leq n_x \leq 127$ at
$t=10^4$ (red/gray) and $10^6$ (black).}
\label{fig6}
\end{figure}

\begin{figure}[ht!]
\begin{center}
\includegraphics[width=8cm]{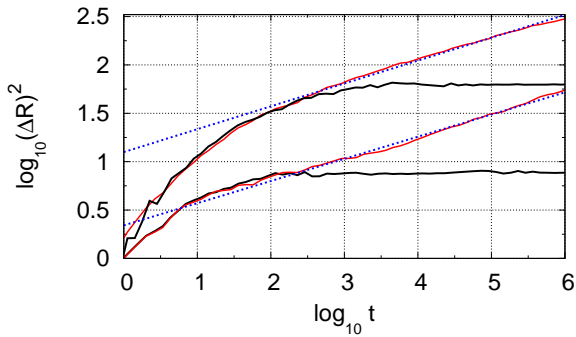} 
\end{center}
\caption{(Color online)
Average of the squared spreading $(\Delta R)^2$ 
as a function of time for two
different values of the disorder strength $W=10$
(top red/gray and black curves) 
 and $W=15$ (bottom  red/gray and black curves)
for $\beta=1$ (red/gray curves) and 
$\beta=0$ (black curves).
The slopes of the straight line fits
for $100 \leq t \leq 10^6$ give  
$\alpha_2=0.236 \pm 0.003$ for $W=10$ 
and $0.229 \pm 0.003$ for $W=15$.
The lattice size is as in Fig.~\ref{fig6}.}
\label{fig7}
\end{figure}

Here we present results for the model (\ref{eq1}) in 2D.
All results are averaged over $N_d=10$ disorder realisations.
The time evolution of the probability distribution
$w_{n_x,n_y}$ is shown in Fig.~\ref{fig6} for $W=10$.
At $\beta=0$ the probability is localized 
while at $\beta=1$ it slowly spreads over the lattice.
The second moment of the space displacement 
$(\Delta R)^2=(\Delta n_x)^2+(\Delta n_y)^2$
as a function of time is shown in Fig.~\ref{fig7}.
The growth is well described by the algebraic dependence
$(\Delta R)^2 = D t^{\alpha_2}$. The fit gives the values of the exponent
$\alpha_2=0.236 \pm 0.003$ for $W=10$ and
$0.229 \pm 0.003$ for  $W=15$. 
Taking into account that the growth of
$(\Delta R)^2$ is rather slow 
and that there are fluctuations related to disorder
averaging the agreement of the exponent
$\alpha_2$ with the theoretical value
$\alpha=1/4$ (\ref{eq3})
can be considered as rather good.
In addition the value of $\alpha_2$ in 2D
is decreased compared to the value  $\alpha_1$ in  1D.
Their ratio $\alpha_2/\alpha_1 = 0.233/0.325 = 0.717$
is rather close to the theoretical value 
$5/8$ given by Eq.(\ref{eq3}). 
According to (\ref{eq3}) the ratio
$D(W=10)/D(W=15) = (l(W=10)/l(W=15))^{1/2} =
((\Delta n(W=10))_0^2/(\Delta n(W=15))_0^2))^{1/4} \approx 1.9$
where $(\Delta n)_0^2$ are the values taken at $\beta=0$.
From the data of Fig.~\ref{fig7} at $\beta=0$ 
we have this ratio to be  1.9 while from data at $\beta=1$
we obtain its value as  5 that can be considered as satisfactory
taking into account all fluctuations.

\begin{figure}[htb!]
\begin{center}
\includegraphics[width=8cm]{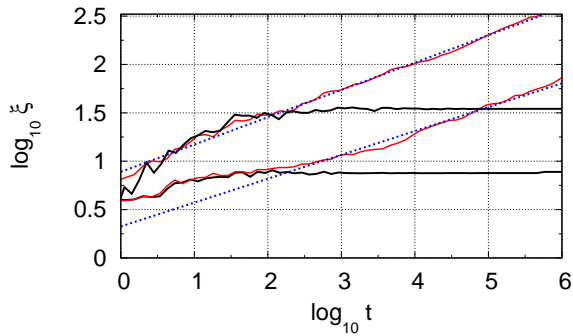} 
\end{center}
\caption{(Color online)
Same as in  Fig.~\ref{fig7} but for the IPR $\xi$. 
The slopes of the straight line fit are
$\nu=0.282\pm 0.002$ for $W=10$ 
(top red/gray and black curves) and 
$\nu=0.247\pm 0.005$ for $W=15$ 
(bottom red/gray and black curves); black curves are for $\beta=0$,
red/gray curves are for $\beta=1$.}
\label{fig8}
\end{figure}

The time dependence of 
the IPR $\xi$ is shown in Fig.~\ref{fig8}.
The fit gives the algebraic growth with the exponents
$\nu= 0.282\pm 0.002$  for $W=10$ and
$\nu=0.247\pm 0.005$ for $W=15$
that is in a good agreement  with the theoretical value
$1/4$ (\ref{eq3}). We note that in 2D 
the exponents $\alpha_2$ and $\nu$ become rather close.
This indicates that multi-fractal effects
become less pronounced in 2D. 

\begin{figure}[htb!]
\begin{center}
\includegraphics[width=8cm]{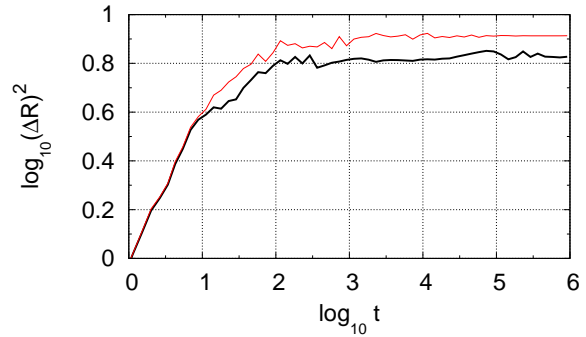} 
\end{center}
\caption{(Color online) 
Average of the squared spreading $(\Delta R)^2$ as a function of time for
for $\beta=0$ (black curve) and $\beta=0.033$ (red/gray curve)
at $W=15$.}
\label{fig9}
\end{figure}

Finally, in Fig.~\ref{fig9} we present the 
comparison of the behaviors of linear system
$\beta=0$ and the one at weak nonlinearity $\beta=0.033$.
These data show that for 
$\beta < \beta_c$ the behaviors of the two systems
are rather similar, for times explored in our numerical simulations,
that is in agreement
with the theoretical expectations described in Section II.
According to our data $\beta_c > 0.033$ in 2D.

To characterize the nonlinear evolution of system (\ref{eq1})
in an additional way we computed another characteristics
which we call nonlinear fidelity defined as
 $f(t)=|\bra{\tilde{\psi}_{n,m}(t)}\ket{\psi_{n,m}}(t)|^2$,
where $\tilde{\psi}_{n,m}(t)$ is a small perturbation
of $\psi_{n,m}$ at $t=0$. For the linear system with $\beta=0$
the fidelity $f(t)$ remains constant during time evolution.
However, for nonlinear dynamics $f(t)$ starts to depend on time.
Indeed, the perturbation changes the nonlinear potential
and the system starts to evolve with a slightly different
effective Hamiltonian that leads to a decrease of fidelity.
Such a behavior reminds the fidelity decay
studied in systems of quantum chaos (see e.g.  review
\cite{prosen}). We note that recently the fidelity decay in the GPE 
has been studied in  \cite{manfredi}, however, there 
the amplitude of random potential was considered as a
very small perturbation while in our case the disordered potential
is strong and plays a dominant role. Also in \cite{manfredi}
the fidelity was considered for perturbation of potential
while we consider the perturbation of nonlinear field
that was  not addressed in \cite{manfredi}.
\begin{figure}[htb!]
\begin{center}
\includegraphics[width=8cm]{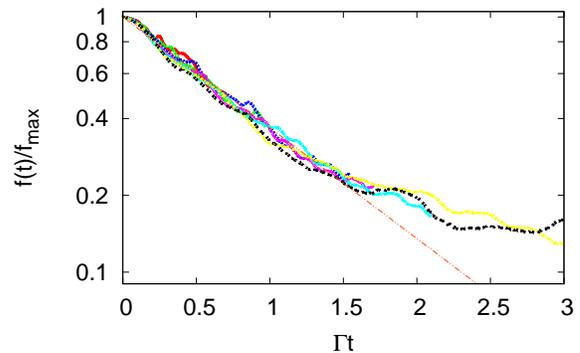} 
\end{center}
\caption{(Color online) Dependence of averaged $f(t)/f_{\rm max}$  
as a function of the rescaled time $\Gamma t$ for 
$W=8$, $\beta=1$ and 
7 different values of perturbed probability $0.001 \leq \delta P \leq 0.03$. 
The average is done over $N_d=12$ disorder realizations.
The straight line shows the dependence
$f(t)/f_{\rm max}=\exp(-\Gamma t)$.}
\label{fig10}
\end{figure}

\begin{figure}[htb!]
\begin{center}
\includegraphics*[width=8cm]{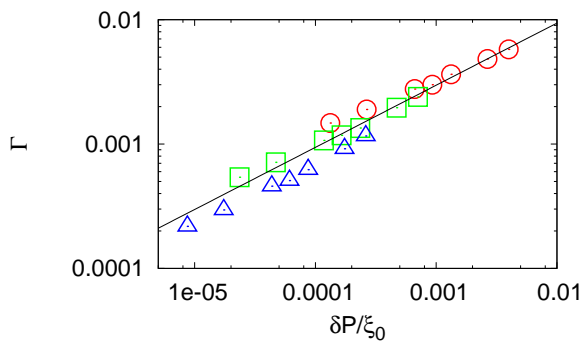} 
\end{center}
\caption{(Color online) Dependence of the fidelity decay rate $\Gamma$
on rescaled perturbation $\delta P/\xi_0$ for
$W=8$ (triangles), $10$ (squares), $15$ (circles).
The straight line  shows the algebraic dependence 
given by fit: $\Gamma = (\delta P/\xi_0)^{\eta}$
with $\eta= 0.486$.}
\label{fig11}
\end{figure}

To study the properties of $f(t)$ we start at $t=0$ from
one lattice state for $\psi$ while the state $\tilde{\psi}$
has a part $\delta P$ of total probability  transferred to
8 nearby lattice sites (e.g. at $t=0$ for $\psi$
the total probability is at a certain lattice site,
while for $\tilde{\psi}$ this site contains the probability $1-\delta P$
and nearby 8 sites have equal probabilities 
$\delta P/8$).
For convenience we fit the decay of fidelity normalized 
by its maximal value
$f_{max}$ by the exponential decay $f(t)/f_{max} = \exp(-\Gamma t)$
with a certain decay rate $\Gamma$. The value of $\Gamma$
is fixed by the condition that curves for various values of $\delta P$
are approximately superimposed on one scaling curve
as it is shown for example in Fig.~\ref{fig10}
for $W=8$ and $\beta=1$. The same procedure was done
for other values of disorder $W$. The resulting dependence 
of $\Gamma$ on $\delta P$ and the average IPR $\xi_0$ at 
large times at $\beta=0$ is shown in Fig.~\ref{fig11}.
The data can be described by the dependence 
$\Gamma \sim (\delta P/\xi_0)^{1/2}$. We interpret
this in the following way: the perturbation$\delta P$ 
spreads over $\xi_0$ states and  gives a
modification of the nonlinear potential 
$\delta |\psi|^2 \sim (\delta P/\xi_0)^{1/2}$
that determines a typical transition frequency
to other states leading to $\Gamma \sim \beta \delta |\psi|^2
\sim \beta (\delta P/\xi_0)^{1/2}$. 
A more detailed check of the functional dependence requires
larger variation of $\xi_0$ that can be done in future
studies.

\begin{figure}[htb!]
\begin{center}
\includegraphics*[width=8cm]{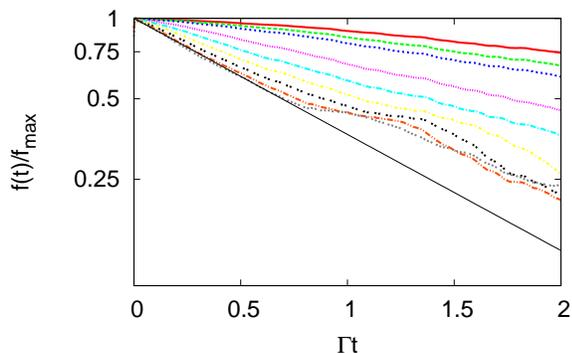} 
\end{center}
\caption{(Color online) 
Rescaled fidelity decay as a function of time, 
for $W=8,\, \beta = 1$ 
and  values of $\delta P$ from 0.005 (top curve)  to 0.7 
(bottom curve); 
averaging is done over $N_d=12$ disorder
realisations. 
At large $\delta P$ the behavior of $f(t)/f_{\rm max}$ 
becomes independent of $\delta P$. The straight line 
shows the decay with the saturated value of $\Gamma_s$:
$f(t)/f_{max}=\exp(-\Gamma_s t)$.}
\label{fig12}
\end{figure}

\begin{figure}[htb!]
\begin{center}
\includegraphics*[width=8cm]{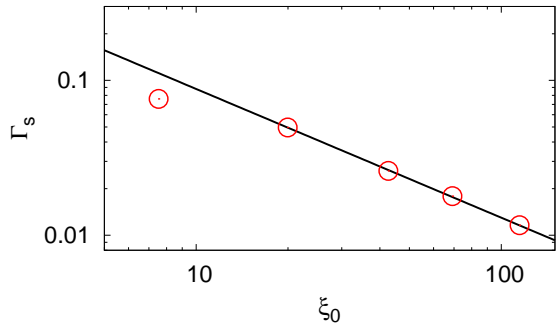} 
\end{center}
\caption{(Color online) Dependence of saturated decay rate $\Gamma_s$ 
(like in Fig.~\ref{fig12}) on 
the IPR $\xi_0$ for $W=15,\,12,\,10,\,9,\,8$
(from left to right). The slope of the straight line fit
is $ -0.831\pm 0.0095 $.}
\label{fig13}
\end{figure}

It is interesting to note that with the increase
of $\delta P$ the growth of the decay rate $\Gamma$
becomes saturated and $\Gamma$ reaches its 
saturated value $\Gamma_s$ (see Fig.~\ref{fig12}).
The dependence of $\Gamma_s$ on $\xi_0$
is shown in Fig.~\ref{fig13}.
Except the strongly localized case at $W=15$
the dependence is satisfactorily described
by $\Gamma_s \sim \beta/\xi_0$. This corresponds to
the situation when the perturbation of nonlinear
field is rather strong and the decay of fidelity
is given by a typical nonlinear frequency shift
$\delta \omega \sim \beta |\psi_n|^2 \sim \beta/\xi_0$.
Of course, this relation is valid on relatively short
time scales used for investigation of fidelity decay
(Figs.~\ref{fig10}-\ref{fig13}) when
$ |\psi_n|^2 \sim \xi_0$. On a larger time scales
the grows of $\xi(t)$ with time should be taken into account.  

The nonlinear fidelity decay gives new additional characteristics 
of nonlinear field evolution on moderate time scales.

\section{Conclusions}
The results of extensive numerical simulations presented above
show that at moderate
nonlinearity  in disordered lattices 
with localized linear eigenmodes  in 1D and 2D
there is an  algebraic spreading
over the lattice with the number of populated sites growing as
$\Delta n \propto t^{\nu}$ (see Eq.~(\ref{eq3})).
This spreading continues up to enormously large times $t=10^9$
while for the linear problem the localization
takes place on a time scale $t_{loc} \sim 10 $ or $100$. 
This result is in a satisfactory agreement with the previous studies
\cite{dls1993,dls2008}. 
The numerical data are obtained on extremely
large time scales (up to $t=10^9$ in dimensionless units)
that are by 4 (Figs.~\ref{fig7},\ref{fig8})
to 8 (Figs.~\ref{fig4},\ref{fig5})  orders of magnitude
larger than the time scale of Anderson localization
(down to $t_{loc}=10$). This indicates that the 
numerical results demonstrate the real asymptotic regime
of algebraic growth. Even if the numerical simulations
do not allow to make rigorous conclusions about
the asymptotic spreading at infinite times we think that
the enormous difference between $t_{loc}$ scale and 
the computational times reached in our numerical simulations
favors the conclusion that
the presented numerical data give the real asymptotic behavior.
The theoretical exponent of spreading $\nu=d/(3d+2)$ 
given by Eq.~(\ref{eq3}) is in a good agreement with the numerical 
data for the KNR model in 1D (\ref{eq5}) and the 2D Anderson model (\ref{eq1})
(see Figs.~\ref{fig4},\ref{fig5},\ref{fig7},\ref{fig8}).
For the 1D Anderson model (\ref{eq1}) the 
numerical value of the exponent $\nu$ shows about 20-30\%
deviation from the theoretical value.
We think that such a deviation should be attributed to
a specific property of the 1D Anderson model
which has no diffusive regime showing a direct
transition from a ballistic dynamics to localization
(see e.g. \cite{kramer1993}).
The data obtained for a weak nonlinearity with $\beta \leq 0.03$
show no spreading up to times $t = 10^6, 10^8$
(see Fig.~\ref{fig9}, Ref.~\cite{dls2008}, Fig.~\ref{fig4}, Fig.~\ref{fig5}
respectively). A small spreading seeing in the KNR model
at enormously large time $t=10^9$
may indicate that a very slow (logarithmic ?)
spreading in time is not completely excluded
and processes like the Arnold diffusion \cite{chirikov} may be present.
However, this spreading, even if present, is so slow that
in global the presented numerical data can be 
considered as a confirmation of the theoretical
expectation according to which 
for a typical initial state the localization is preserved at 
$\beta < \beta_c$. Our data indicate that $\beta_c \sim 1/30$.
Indeed, the spreading behavior is qualitatively different
for $\beta \sim1$ and $\beta \sim 1/30$.

It is possible that such type of slow
probability and energy spreading over disordered
lattices may play an important role in
complex molecules giving more rapid 
propagation of probability and energy 
along molecular chains compared to a simple diffusion produced by noise.
It would be interesting to observe the nonlinear destruction of
localization for BEC in disordered potential or for
nonlinear waves in photonic lattices but this is rather hard
task since very long observation times are required for that.

We thank A.S.Pikovsky for stimulating discussions,
one of us (DLS) thanks the participants of the NLSE workshop
at the Lewiner Institute at the Technion for useful discussions.

{\it Note added:} after the submission of this paper there 
appeared the preprint \cite{flacharx} where 
the same nonlinear 1D Anderson model
is investigated numerically and analytically.



\begin{thebibliography}{99}
\bibitem{anderson1958} P.W.~Anderson, Phys. Rev. {\bf 109}, 1492 (1958).
\bibitem{montambaux} E.~Akkermans and G.~Montambaux,
        {\it Mesoscopic Physics of Electrons and Photons},
         Cambridge Univ. Press, Cambridge (2007).
\bibitem{pitaevskii} F.~Dalfovo, S.~Giorgini, L.P.~Pitaevskii, and
         S.~Strigani, Rev. Mod. Phys. {\bf 71}, 463 (1999).
\bibitem{oberthaler} O.~Morsch and M.~Oberthaler, Rev. Mod. Phys.
                  {\bf 78}, 179 (2006).
\bibitem{inguscio2008} L.~Fallani, C.Fort, and M.Inguscio,
        arxiv:[0804.2888[cond-mat] (2008).
\bibitem{inguscio2005} J.E.~Lye, L.~Fallani, M.~Modugno, D.S.~Wiersma, 
        C.~Fort, and M.~Inguscio, Phys. Rev. Lett. {\bf 95}, 070401 (2005).
\bibitem{aspect2005} D.~Cl\'ement, A.F.~Var\'on, 
        M.~Hugbart, J.A.~Retter, P.~Bouyer,
        L.~Sanchez-Palencia, D.M.Gangardt, G.V.Shlyapnikov, and A.~Aspect,
        Phys. Rev. Lett. {\bf 95}, 170409 (2005).
\bibitem{inguscio12005} C.~Fort, L.~Fallani, V.~Guarrera, J.E.~Lye, M.~Modugno,
        D.S.~Wiersma, and M.~Inguscio, Phys. Rev. Lett. 
        {\bf 95}, 170410 (2005).
\bibitem{ertmer2005} T.Schulte, S.~Drenkelforth, J.Kruse, W.Ertmer, J.~Arlt,
        K.~Sacha, J.Zakrzewski, and M.~Lewenstein,
        Phys. Rev. Lett. {\bf 95}, 170411 (2005).
\bibitem{aspect2006} D.~Cl\'ement, A.F.~Var\'on,  
        J.A.~Retter, L.~Sanchez-Palencia,
        A.~Aspect, and P.~Bouyer, New J. Phys. {\bf 8}, 165 (2006).
\bibitem{phillips2006} C.~Ryu, M.F.~Andersen, 
        A.~Vaziri, M.B.~d'Arcy, J.M.~Grossman,
        K.~Helmerson, and W.D.~Phillips, 
        Phys. Rev. Lett. {\bf 96}, 160403 (2006).
\bibitem{summy2006} G.~Behinaein, V.~Ramareddy, P.~Ahmadi, and G.S.~Summy,
        Phys. Rev. Lett. {\bf 97}, 244101 (2006).
\bibitem{steinberg} J.~F.~Kanem, S.~Maneshi, M.~Partlow, M.~Spanner and
              A.~M.~Steinberg, Phys. Rev. Lett. {\bf 98}, 083004 (2007).
\bibitem{chirikov} B.~V.~Chirikov, Phys. Rep. {\bf 52}, 263 (1979);
                   B.~V.~Chirikov, F.~M.~Izrailev and D.~L.~Shepelyansky,
                  Sov. Scient. Rev. C {\bf 2}, 209 (1981);
                  Physica {\bf 33D}, 77 (1988);
                  B.~Chirikov and D.~Shepelyansky, 
                  Scholarpedia 3(3):3550 (2008).
\bibitem{segev2007} T.~Schwartz, G.~Bartal, S.~Fishman, and M.~Segev,
        Nature {\bf 446}, 52 (2007).
\bibitem{silberberg2008} Y.~Lahini, A.~Avidan, F.~Pozzi, 
        M.~Sorel, R.~Morandotti,
        D.N.~Christodoulides, and Y.~Silberberg, Phys. Rev. Lett. 
        {\bf 100}, 013906 (2008).
\bibitem{cao} H.~Cao, Waves in Random Media {\bf 13}, R1 (2003).
\bibitem{kivshar} S.A.~Gredeskul and Y.S.Kivshar, 
        Phys. Rep. {\bf 216}, 1 (1992).
\bibitem{pavloff1} T.~Paul, P.~Schlagheck, P.~Leboef, and N.Pavloff,
         Phys. Rev. Lett. {\bf 98}, 210602 (2007).
\bibitem{dls1991} F.~Benvenuto, G.~Casati, A.S.Pikovsky, and D.L.Shepelyansky,
        Phys. Rev. A {\bf 44}, R3423 (1991).
\bibitem{dls1993} D.L.~Shepelyansky, Phys. Rev. Lett. {\bf 70}, 1787 (1993).
\bibitem{pavloff} N.~Bilas and N.~Pavloff, Phys. Rev. Lett. 
        {\bf 95}, 130403 (2005).
\bibitem{aspect2007} L.~Sanchez-Palencia,  D.~Cl\'ement, P.~Lugan, P.~Bouyer,
         G.V.Shlyapnikov, and A.~Aspect,  
         Phys. Rev. Lett. {\bf 98}, 210401 (2007).
\bibitem{flach2008} G.~Kopidakis, S.~Komineas, 
        S.~Flach, and S.~Aubry, Phys. Rev. Lett.
        {\bf 100}, 084103 (2008).
\bibitem{dls2008} A.S.~Pikovsky and D.L.~Shepelyansky,
        Phys. Rev. Lett. {\bf 100}, 094101 (2008).
\bibitem{lebowitz} A.~Dhar and J.L.~Lebowitz,
         Phys. Rev. Lett. {\bf 100}, 134301 (2008).
\bibitem{skipetrov2000} S.E.~Skipetrov, and 
        R.~Maynard, Phys. Rev. Lett. {\bf 85}, 736 (2000).
\bibitem{shapiro2007} B.~Shapiro, Phys. Rev. Lett. {\bf 99}, 060602 (2007).
\bibitem{wellens} T.~Wellens and B.~Gremaud,
        Phys. Rev. Lett. {\bf 100}, 033902 (2008).
\bibitem{shapiro2008} S.E.~Skipetrov, A.~Minguzzi, B.A. van Tiggelen,
        and B.~B.Shapiro, Phys. Rev. Lett. {\bf 100}, 165301 (2008).
\bibitem{frolich} J.~Frolich, T.~Spencer, and C.E.Wayne,
        J. Stat. Phys. {\bf 42}, 247 (1986).
\bibitem{james2005} G.~Iooss, and G.~James, Chaos {\bf 15}, 015113 (2005).
\bibitem{zhirov} O.V.~Zhirov, G.~Casati, and D.L.~Shepelyansky,
        Phys. Rev. E {\bf 65}, 026220 (2002);
        arXiv:cond-mat/0501188 (2005).
\bibitem{soffer} S.~Fishman, Y.Krivopalov, and A.~Soffer,
        J. Stat. Phys. {bf 131}, 843 (2008).
\bibitem{wang1} W.-M.~Wang and Z.Zhang,
        preprint arXiv:0805.3520[math-ph]  (2008).
\bibitem{wang2} J.~Bourgain and W.-M.~Wang,
        J. Eur. Math. Soc. {\bf 10}, 1 (2008);
        preprint arXiv:0805.4632 [math.DS]  (2008).
\bibitem{kramer1993} B.~Kramer, and A.~MacKinnon, 
        Rep. Prog. Phys. {\bf 56}, 1469 (1993).
\bibitem{dls1994} D.L.~Shepelyansky, Phys. Rev. Lett.
        {\bf 73}, 2607 (1994).
\bibitem{dls1995} P.~Jacquod and D.L.~Shepelyansky,
        Phys. Rev. Lett. {\bf 75}, 3501 (1995).
\bibitem{chirikov1} B.V.~Chirikov and D.L.~Shepelyanskii,
        Sov. J. Nucl. Fiz. {\bf 36}, 908 (1982)
        [Yader. Fiz. {\bf 36}, 1563 (1982)].
\bibitem{fishman}S.~Fishman, D.~R.~Grempel, and R.~E.~Prange, 
        Phys. Rev. Lett. {\bf 49}, 509 (1982).        
\bibitem{prosen} T.~Gorin, T.~Prosen, T.~H.~Seligman and M.~Znidaric,
        Phys. Rep. {\bf 435}, 33 (2006).
\bibitem{manfredi} G.~Manfredi and P.-A.Hervieux,
        Phys. Rev. Lett. {\bf 100}, 050405 (2008).
\bibitem{flacharx} S.~Flach, D.O.~Krimer, and Ch.~Skokos,
        preprint arXiv:0805.4693[cond-mat] (2008).
\end{thebibliography}
\end{document}